\documentclass[pra,onecolumn]{revtex4}%
\usepackage{amsfonts}
\usepackage{amsmath}
\usepackage{amssymb}
\usepackage{graphicx}%
\providecommand{\U}[1]{\protect\rule{.1in}{.1in}}

\begin{document}
\title{Generation of atomic entangled states in a bi-mode cavity via adiabatic passage}
\author{Li-Bo Chen}
\author{Peng Shi}
\author{Yong-Jian Gu}
\email{yjgu@ouc.edu.cn}
\author{Lin Xie}
\author{Li-Zhen Ma}
\affiliation{Department of Physics, Ocean University of China, Qingdao 266100, People's
Republic of China}

\begin{abstract}
We propose schemes to prepare atomic entangled states in a bi-mode cavity via
stimulated Raman adiabatic passage (STIRAP) and fractional stimulated Raman
adiabatic passage (f-STIRAP) techniques. According to the simulation results,
our schemes keep the cavity modes almost unexcited and the atomic excited
states are nearly unpopulated during the whole process. The simulation also
shows that the error probability is very small.

\end{abstract}

\maketitle

\section{Introduction}

Entanglement, especially high dimensional entanglement, has been recognized as
the crucial ingredient in many quantum information processes, such as quantum
communication \cite{bennett,bennett2,Riedmatten,Zukowski,Molina-Terriza},
quantum computation \cite{Ralph,G Molina,Cabello}. Thus more and more
attention has been paid to generating qubit or qudit entangled states in
cavity QED \cite{Francica,Delgado,zou,zheng1,zheng2,zheng3,Lin1,G.W.
Lin,Asoka}, ion trap \cite{X. W. Wang,Linington,X.H. Huang}, superconducting
qubit \cite{Jian Li,Fay,Vitali,G. Chen}, linear optical system \cite{G.
Vallone,F. W. Sun}, semiconductor quantum dot \cite{G. E. Murgida,S. K.
Saikin}, and so on.

The technique of stimulated Raman adiabatic passage (STIRAP), which was shown
theoretically \cite{Oreg} and implemented experimentally in population
transfer process in molecules and atoms \cite{Gaubatz,Gaubatz2}, has been
broadly used in quantum information processing (QIP)
\cite{Kral,ber,kis,go,la,moll,mh,ye1,ye2,songjie,sham}. In this technique are
involved two delayed, but partly overlapping pulses -- pump and Stokes -- with
the Stokes pulse applied first. L. Wang \textit{et al} \cite{L.wang}
experimentally and theoretically demonstrate that the atomic coherence can be
completely transferred or arbitrarily distributed among the different levels
in a four-level atomic (tripod) scheme by STIRAP. An extension of STIRAP,
called fractional stimulated Raman adiabatic passage (f-STIRAP)
\cite{nv,ma,Klein,l.b.chen}, allows the creation of a coherent superposition
of two ground states in $\Lambda$-type system. Different from STIRAP, the
f-STIRAP requires that the two pulses vanish at a constant finite ratio of
amplitudes. Wang \textit{et al} \cite{wang} have demonstrated storage and
selective release of a light pulse in a Pr:YSO crystal, which is based on
atomic spin coherence created by the f-STIRAP.

In this Brief Report, we present schemes for generating entangled states of
two qubits and two qutrits via STIRAP or f-STIRAP techniques. Two $^{87}Rb$
atoms are trapped in an optical cavity and drived by two different lasers. By
choosing appropriate parameters we can create entangled atomic states.
According to the simulation results, our schemes keep the cavity modes almost
unexcited and the atomic excited states are nearly unpopulated during the
whole process. The simulation also shows that the error probability is very small.

\section{The fundamental model}

\begin{figure}[ptb]
\centering
\includegraphics[scale=0.8, angle=0]{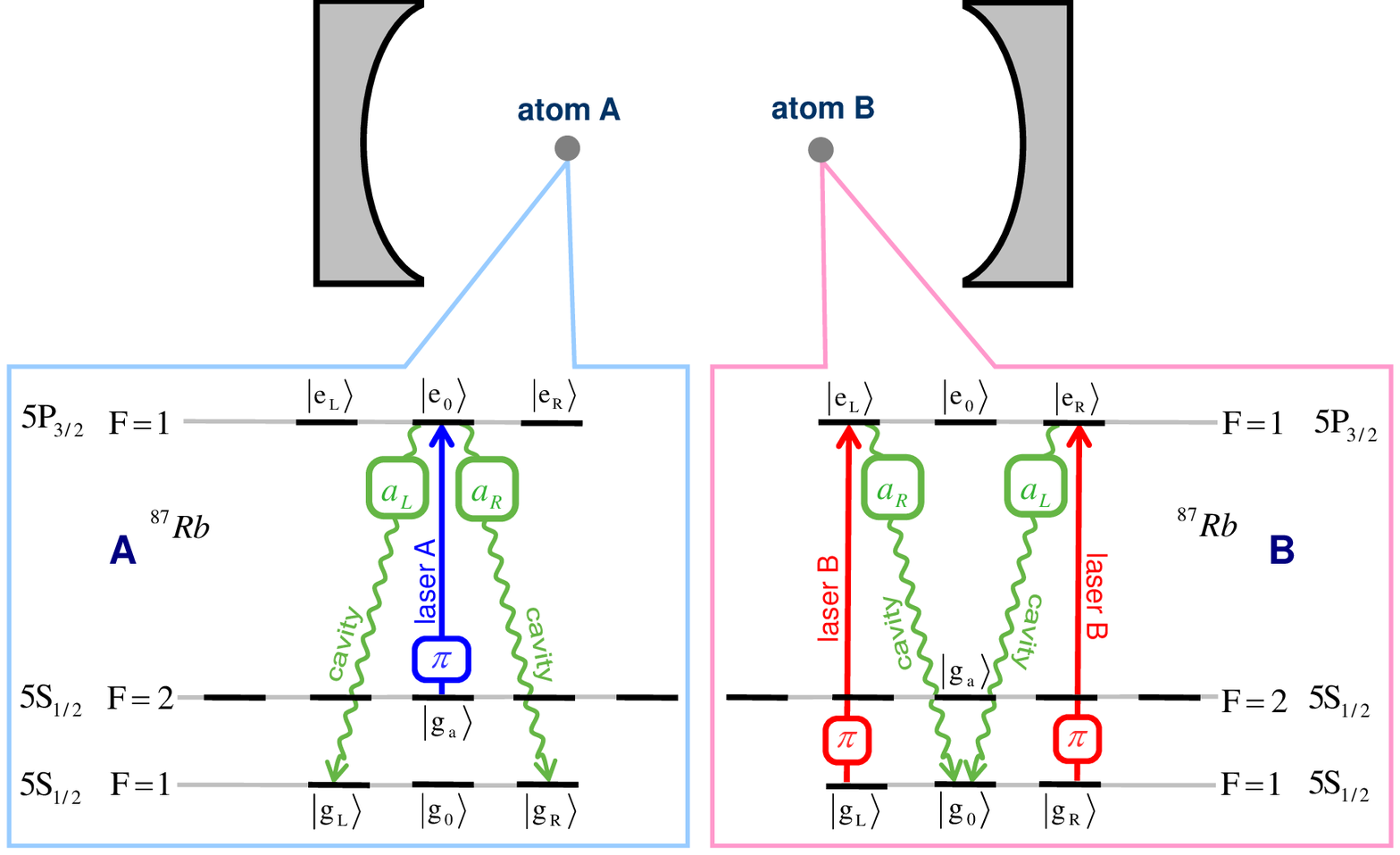}\caption{Two $^{87}Rb$ atoms are
trapped in a bi-mode cavity. The states $\left\vert g_{L}\right\rangle $
($\left\vert e_{L}\right\rangle $), $\left\vert g_{0}\right\rangle $
($\left\vert e_{0}\right\rangle $), and $\left\vert g_{R}\right\rangle $
($\left\vert e_{R}\right\rangle $) correspond to the $^{87}Rb$\ atom hyperfine
levels $F=1$ of $5S_{1/2}$ ($5P_{3/2}$), while $\left\vert g_{a}\right\rangle
$ corresponds to $F=2$, $m_{F}=0$ of $5S_{1/2}$. The atom $A$'s transition
$\left\vert g_{a}\right\rangle _{A}\leftrightarrow\left\vert e_{0}%
\right\rangle _{A}$ is driven resonantly by a $\pi$-polarized classical field
with Rabi frequency $\Omega_{A}$; $\left\vert e_{0}\right\rangle
_{A}\leftrightarrow\left\vert g_{R}\right\rangle _{A}$ $\left(  \left\vert
e_{0}\right\rangle _{A}\leftrightarrow\left\vert g_{L}\right\rangle
_{A}\right)  $ is resonantly coupled to the cavity mode $a_{R}$ $\left(
a_{L}\right)  $ with coupling constant $g_{A}$. The atom $B$'s transition
$\left\vert g_{L}\right\rangle _{B}\leftrightarrow\left\vert e_{L}%
\right\rangle _{B}$ $\left(  \left\vert g_{R}\right\rangle _{B}\leftrightarrow
\left\vert e_{R}\right\rangle _{B}\right)  $\ is driven resonantly by a $\pi
$-polarized classical field with Rabi frequency $\Omega_{B}$; $\left\vert
e_{L}\right\rangle _{B}\leftrightarrow\left\vert g_{0}\right\rangle _{B}$
$\left(  \left\vert e_{R}\right\rangle _{B}\leftrightarrow\left\vert
g_{0}\right\rangle _{B}\right)  $ is resonantly coupled to the cavity mode
$a_{R}$ $\left(  a_{L}\right)  $ with coupling constant $g_{B}$.}%
\end{figure}We consider the situation described in Figure 1, where two atoms
are trapped in a bi-mode optical cavity. The relevant atomic levels and
transitions are also depicted in this figure; such level structures can be
achieved in $^{87}Rb$ \cite{wilk,Weber}. The states $\left\vert g_{L}%
\right\rangle $, $\left\vert g_{0}\right\rangle $, $\left\vert g_{R}%
\right\rangle $ and $\left\vert g_{a}\right\rangle $ correspond to $^{87}%
Rb$\ atom hyperfine levels $\left\vert F=1\text{, }m_{F}=-1\right\rangle $,
$\left\vert F=1\text{, }m_{F}=0\right\rangle $, $\left\vert F=1\text{, }%
m_{F}=1\right\rangle $ of $5S_{1/2}$ and $\left\vert F=2\text{, }%
m_{F}=0\right\rangle $ of $5S_{1/2}$, while $\left\vert e_{L}\right\rangle $,
$\left\vert e_{0}\right\rangle $ and $\left\vert e_{R}\right\rangle $
correspond to $\left\vert F=1\text{, }m_{F}=-1\right\rangle $, $\left\vert
F=1\text{, }m_{F}=0\right\rangle $ and $\left\vert F=1\text{, }m_{F}%
=1\right\rangle $ of $5P_{3/2}$. Initially, the atoms $A$ and $B$ are prepared
in the state $\left\vert g_{a}\right\rangle _{A}$ and $\left\vert
g_{0}\right\rangle _{B}$ respectively, and the cavity mode is in the vacuum
state. The atom $A$'s transition $\left\vert g_{a}\right\rangle _{A}%
\leftrightarrow\left\vert e_{0}\right\rangle _{A}$ is driven resonantly by a
$\pi$-polarized classical field with Rabi frequency $\Omega_{A}$; $\left\vert
e_{0}\right\rangle _{A}\leftrightarrow\left\vert g_{R}\right\rangle _{A}$
$\left(  \left\vert e_{0}\right\rangle _{A}\leftrightarrow\left\vert
g_{L}\right\rangle _{A}\right)  $ is resonantly coupled to the cavity mode
$a_{R}$ $\left(  a_{L}\right)  $ with coupling constant $g_{A}$. The atom
$B$'s transition $\left\vert g_{L}\right\rangle _{B}\leftrightarrow\left\vert
e_{L}\right\rangle _{B}$ $\left(  \left\vert g_{R}\right\rangle _{B}%
\leftrightarrow\left\vert e_{R}\right\rangle _{B}\right)  $\ is driven
resonantly by a $\pi$-polarized classical field with Rabi frequency
$\Omega_{B}$; $\left\vert e_{L}\right\rangle _{B}\leftrightarrow\left\vert
g_{0}\right\rangle _{B}$ $\left(  \left\vert e_{R}\right\rangle _{B}%
\leftrightarrow\left\vert g_{0}\right\rangle _{B}\right)  $ is resonantly
coupled to the cavity mode $a_{R}$ $\left(  a_{L}\right)  $ with coupling
constant $g_{B}$. In the rotating wave approximation, the interaction
Hamiltonian can be written as (setting $\hbar=1$)
\begin{align}
H_{0} &  =g_{A}(t)a_{R}\left\vert e_{0}\right\rangle _{A}\left\langle
g_{R}\right\vert +g_{A}(t)a_{L}\left\vert e_{0}\right\rangle _{A}\left\langle
g_{L}\right\vert +\Omega_{A}(t)\left\vert e_{0}\right\rangle _{A}\left\langle
g_{a}\right\vert \nonumber\\
&  +g_{B}(t)a_{R}\left\vert e_{L}\right\rangle _{B}\left\langle g_{0}%
\right\vert +g_{B}(t)a_{L}\left\vert e_{R}\right\rangle _{B}\left\langle
g_{0}\right\vert +\Omega_{B}(t)\left\vert e_{L}\right\rangle _{B}\left\langle
g_{L}\right\vert +\Omega_{B}(t)\left\vert e_{R}\right\rangle _{B}\left\langle
g_{R}\right\vert +H.c..\label{1}%
\end{align}
In the following we write the state of the system as $\left\vert
A,B,n_{R},n_{L}\right\rangle $, where $A$ and $B$ denote the states of the
atoms, and $n_{L.R}$ the number of $L$ or $R$ polarized photons of the cavity.
The subspace $S$ spanned by states $\{\left\vert g_{a},g_{0},0,0\right\rangle
$, $\left\vert e_{0},g_{0},0,0\right\rangle $, $\left\vert g_{L}%
,g_{0},0,1\right\rangle $, $\left\vert g_{L},e_{R},0,0\right\rangle $,
$\left\vert g_{L},g_{R},0,0\right\rangle $, $\left\vert g_{R},g_{0}%
,1,0\right\rangle $, $\left\vert g_{R},e_{L},0,0\right\rangle $, $\left\vert
g_{R},g_{L},0,0\right\rangle \}$ is an 8-dimensionnal invariant subspace of
the Hamiltonian (1). It can be verified that in the subspace $S$, the
Hamiltonian has the following dark state:
\begin{align}
\left\vert D(t)\right\rangle  &  =C[2g_{A}(t)\Omega_{B}(t)\left\vert
g_{a},g_{0},0,0\right\rangle -\Omega_{A}(t)\Omega_{B}(t)\left(  \left\vert
g_{L},g_{0},0,1\right\rangle +\left\vert g_{R},g_{0},1,0\right\rangle \right)
\nonumber\\
&  +g_{B}(t)\Omega_{A}(t)\left(  \left\vert g_{L},g_{R},0,0\right\rangle
+\left\vert g_{R},g_{L},0,0\right\rangle \right)  ],\label{2}%
\end{align}
where we assume $g_{i}$, $\Omega_{i}$ are real, and $C^{-2}=4g_{A}^{2}%
\Omega_{B}^{2}+2\Omega_{A}^{2}\Omega_{B}^{2}+2g_{B}^{2}\Omega_{A}^{2}$. Under
the condition
\begin{equation}
g_{A}(t),g_{B}(t)\gg\Omega_{A}(t),\Omega_{B}(t),\label{3}%
\end{equation}
we have%
\begin{equation}
\left\vert D(t)\right\rangle \sim2g_{A}(t)\Omega_{B}(t)\left\vert g_{a}%
,g_{0},0,0\right\rangle +g_{B}(t)\Omega_{A}(t)\left\vert g_{L},g_{R}%
,0,0\right\rangle +g_{B}(t)\Omega_{A}(t)\left\vert g_{R},g_{L}%
,0,0\right\rangle .\label{4}%
\end{equation}

\section{Generation \ of a two-qubit entangled state via STIRAP}

\bigskip Suppose the initial state of the system is $\left\vert g_{a}%
,g_{0},0,0\right\rangle $, if we design pulse shapes and sequence such that
\begin{align}
\lim_{t\rightarrow-\infty}\frac{g_{B}(t)\Omega_{A}(t)}{g_{A}(t)\Omega_{B}(t)}
&  =0,\label{5}\\
\lim_{t\rightarrow+\infty}\frac{g_{A}(t)\Omega_{B}(t)}{g_{B}(t)\Omega_{A}(t)}
&  =0, \label{6}%
\end{align}
it follows from Eq. (4) that we can adiabatically transfer the initial state
$\left\vert g_{a},g_{0},0,0\right\rangle $ to a superposition of $\left\vert
g_{L},g_{R},0,0\right\rangle $\ and $\left\vert g_{R},g_{L},0,0\right\rangle
$, i.e., $1/\sqrt{2}\left(  \left\vert g_{L},g_{R},0,0\right\rangle
+\left\vert g_{R},g_{L},0,0\right\rangle \right)  =1/\sqrt{2}(\left\vert
g_{L},g_{R}\right\rangle +\left\vert g_{R},g_{L}\right\rangle )\left\vert
00\right\rangle _{c}$, which is a product state of the two-atom entangled
state and the cavity mode vacuum state.

The pulse shapes and sequence can be designed by an appropriate choice of the
parameters. The coupling rates are chosen such that $g_{A}(t)=g_{B}(t)=g$,
laser Rabi frequencies are chosen as $\Omega_{A}(t)=\Omega_{0}\exp\left[
-1/200\left(  t-t_{0}\right)  ^{2}/\tau^{2}\right]  $ and $\Omega
_{B}(t)=\Omega_{0}\exp\left[  -1/200t^{2}/\tau^{2}\right]  $, with
$t_{0}=2\tau$ being the delay between pulses \cite{Kral}. With this choice,
conditions (5) and (6) are satisfied.

To evaluate the performance of our scheme, we now consider the dissipative
processes due to spontaneous decay of the atoms from the excited states and
the decay of cavity. The evolution of the system is governed by the
non-Hermitian Hamiltonian%

\begin{equation}
H_{nh}=H_{0}-i\kappa(a_{L}^{+}a_{L}+a_{R}^{+}a_{R})-\gamma\sum_{j=A,B}%
(\left\vert e_{R}\right\rangle _{j}\left\langle e_{R}\right\vert +\left\vert
e_{L}\right\rangle _{j}\left\langle e_{L}\right\vert +\left\vert
e_{0}\right\rangle _{j}\left\langle e_{0}\right\vert ), \label{7}%
\end{equation}
where $\kappa$ is the cavity decay rate and $\gamma$\ is the atomic
spontaneous emission rate. Here we assume that three excited states
$\left\vert e_{R}\right\rangle $,$\ \left\vert e_{L}\right\rangle $, and
$\left\vert e_{0}\right\rangle $ have the same spontaneous emission rate
$\gamma$, and the two cavity modes possess the same loss rate $\kappa$. Figure
2 shows the simulation results of the two-qubit entanglement generation
process, where we choose $g=5\Omega_{0}$, $\tau=$ $\Omega_{0}^{-1}$, the
cavity decay rate\ and the atomic spontaneous emission rate $\kappa
=\gamma=0.005g$ \cite{Spillane,Birnbaum}. The Rabi frequencies of $\Omega
_{A}(t)$, $\Omega_{B}(t)$ are shown in Fig. 2(a). Figure 2(b) shows the time
evolution of populations, in which $P_{1}$, $P_{5}$, and $P_{8}$ denote the
populations of the states $\left\vert g_{a},g_{0},0,0\right\rangle $,
$\left\vert g_{L},g_{R},0,0\right\rangle $, and $\left\vert g_{R}%
,g_{L},0,0\right\rangle $, and the populations of the states $\left\{
\left\vert e_{0},g_{0},0,0\right\rangle \text{, }\left\vert g_{L}%
,g_{0},0,1\right\rangle \text{, }\left\vert g_{L},e_{R},0,0\right\rangle
\text{, }\left\vert g_{R},g_{0},1,0\right\rangle \text{, }\left\vert
g_{R},e_{L},0,0\right\rangle \right\}  $ are almost zero during the whole
dynamics. Finally $P_{5}$ and $P_{8}$ arrive at $0.5$ and $P_{1}$ approaches
0, which means the successful generation of the two qubit entangled state.
Figure 2(c) shows the error probability defined by \cite{goto}:
\begin{equation}
P_{e}=1-\left\vert \left\langle D\left(  t\right)  \right\vert \varphi
_{s}\left(  t\right)  \rangle\right\vert ^{2}, \label{8}%
\end{equation}
here $\left\vert \varphi_{s}\left(  t\right)  \right\rangle $ is the state
obtained by the simulation and $\left\vert D(t)\right\rangle $ is the dark
state defined by Eq. (2). In Figure 2(d), the probability $P_{p}$ with which
one photon appears in the cavity is shown. Figure 2(e) shows the population
$P_{ea}$ of the atoms in excited state $\left\vert e_{i}\right\rangle $
$\left(  i=L\text{, }0\text{, }R\right)  $. From Figures 2(c)-2(e) we conclude
that we can prepare the two qubit entangled state with high success
probability. \begin{figure}[ptb]
\centering
\includegraphics[scale=1.0, angle=0]{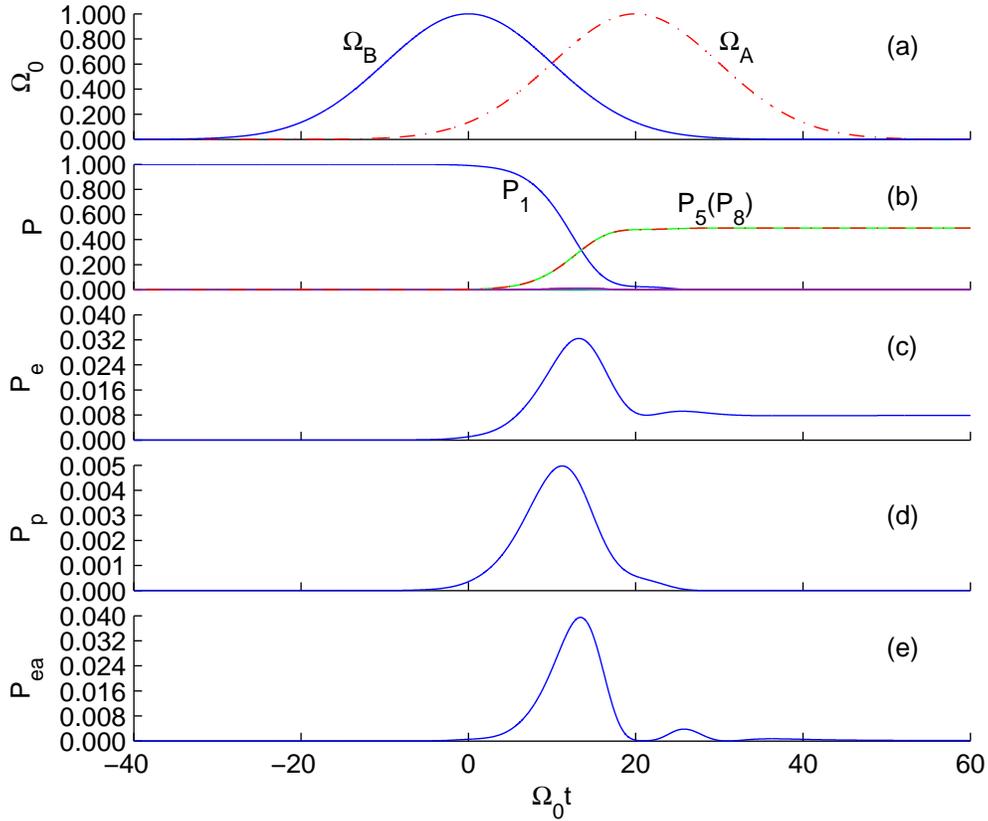}\caption{The simulation results
of the two-qubit entanglement generation process, where we choose
$g=5\Omega_{0}$, $\tau=$ $\Omega_{0}^{-1}$, the cavity decay rate\ and the
atomic spontaneous emission rate $\kappa=\gamma=0.005g$. Figure 2(a): the Rabi
frequency of $\Omega_{A}(t)$, $\Omega_{B}(t)$. Figure 2(b): the time evolution
of populations, in which $P_{1}$, $P_{5}$, and $P_{8}$ denote the populations
of the states $\left\vert g_{a},g_{0},0,0\right\rangle $, $\left\vert
g_{L},g_{R},0,0\right\rangle $, and $\left\vert g_{R},g_{L},0,0\right\rangle
$, and the populations of the states $\left\{  \left\vert e_{0},g_{0}%
,0,0\right\rangle \text{, }\left\vert g_{L},g_{0},0,1\right\rangle \text{,
}\left\vert g_{L},e_{R},0,0\right\rangle \text{, }\left\vert g_{R}%
,g_{0},1,0\right\rangle \text{, }\left\vert g_{R},e_{L},0,0\right\rangle
\right\}  $ are almost zero during the whole dynamics. Figure 2(c): error
probability $P_{e}\left(  t\right)  $ defined by Eq. (8). Figure 2(d): the
probability $P_{p}$ with which one photon appears in the cavity. Figure 2(e):
the population $P_{ea}$ of the atoms in excited state $\left\vert
e_{i}\right\rangle $ $\left(  i=L\text{, }0\text{, }R\right)  $.}%
\end{figure}

\section{Generation of a two-qutrit entangled state via f-STIRAP}

Suppose the initial state of the system is $\left\vert g_{a},g_{0}%
,0,0\right\rangle $, if we design pulse shapes and sequence such that
\begin{align}
\lim_{t\rightarrow-\infty}\frac{g_{B}(t)\Omega_{A}(t)}{g_{A}(t)\Omega_{B}(t)}
&  =0,\\
\lim_{t\rightarrow+\infty}\frac{g_{A}(t)\Omega_{B}(t)}{g_{A}(t)\Omega_{B}(t)}
&  =\frac{1}{2},
\end{align}
the system will end up in the state $1/\sqrt{3}\left(  \left\vert g_{a}%
,g_{0},0,0\right\rangle +\left\vert g_{L},g_{R},0,0\right\rangle +\left\vert
g_{R},g_{L},0,0\right\rangle \right)  =1/\sqrt{3}(\left\vert g_{a}%
,g_{0}\right\rangle +\left\vert g_{L},g_{R}\right\rangle +\left\vert
g_{R},g_{L}\right\rangle )\left\vert 00\right\rangle _{c}$, which is a product
state of the three-dimensional entangled state of the two atoms and the cavity
mode vacuum state. We choose the pulses $g_{A}(t)$, $\Omega_{A}(t)$,
$g_{B}(t)$, and $\Omega_{B}(t)$ the same as in Sec. III, i.e. $g_{A}%
(t)=g_{B}(t)=g$, $\Omega_{A}(t)=\Omega_{0}\exp\left[  -1/200\left(
t-t_{0}\right)  ^{2}/\tau^{2}\right]  +\frac{\Omega_{0}}{2}\exp\left[
-1/200t^{2}/\tau^{2}\right]  $ and $\Omega_{B}(t)=\Omega_{0}\exp\left[
-1/200t^{2}/\tau^{2}\right]  $, with $t_{0}=2\tau$.

Figure 3 shows the simulation results of the two-qutrit entanglement
generation process, where we choose $g$, $\Omega_{0}$, $\tau$, $\kappa$, and
$\gamma$ the same as in Figure 2. The Rabi frequencies of $\Omega_{A}(t)$,
$\Omega_{B}(t)$ are shown in Fig. 3(a). Figure 3(b) shows the time evolution
of populations, in which $P_{1}$, $P_{5}$, and $P_{8}$ denote the populations
of the states $\left\vert g_{a},g_{0},0,0\right\rangle $, $\left\vert
g_{L},g_{R},0,0\right\rangle $, and $\left\vert g_{R},g_{L},0,0\right\rangle
$, and the populations of the states $\left\{  \left\vert e_{0},g_{0}%
,0,0\right\rangle \text{, }\left\vert g_{L},g_{0},0,1\right\rangle \text{,
}\left\vert g_{L},e_{R},0,0\right\rangle \text{, }\left\vert g_{R}%
,g_{0},1,0\right\rangle \text{, }\left\vert g_{R},e_{L},0,0\right\rangle
\right\}  $ are almost zero during the whole dynamics. Finally $P_{1}$,
$P_{5}$, and $P_{8}$ arrive at $1/3$, which means the successful generation of
the two qutrit entangled state. Figure 3(c) shows the error probability during
the process. In Figure 3(d), the probability $P_{p}$ with which one photon
appears in the cavity is shown. Figure 3(e) shows the population $P_{ea}$ of
the atoms in excited state $\left\vert e_{i}\right\rangle $ $\left(
i=L\text{, }0\text{, }R\right)  $. From Figures 3(c)-3(e) we conclude that we
can prepare the two qutrit entangled state with high success probability.
\begin{figure}[ptb]
\centering
\includegraphics[scale=1.0, angle=0]{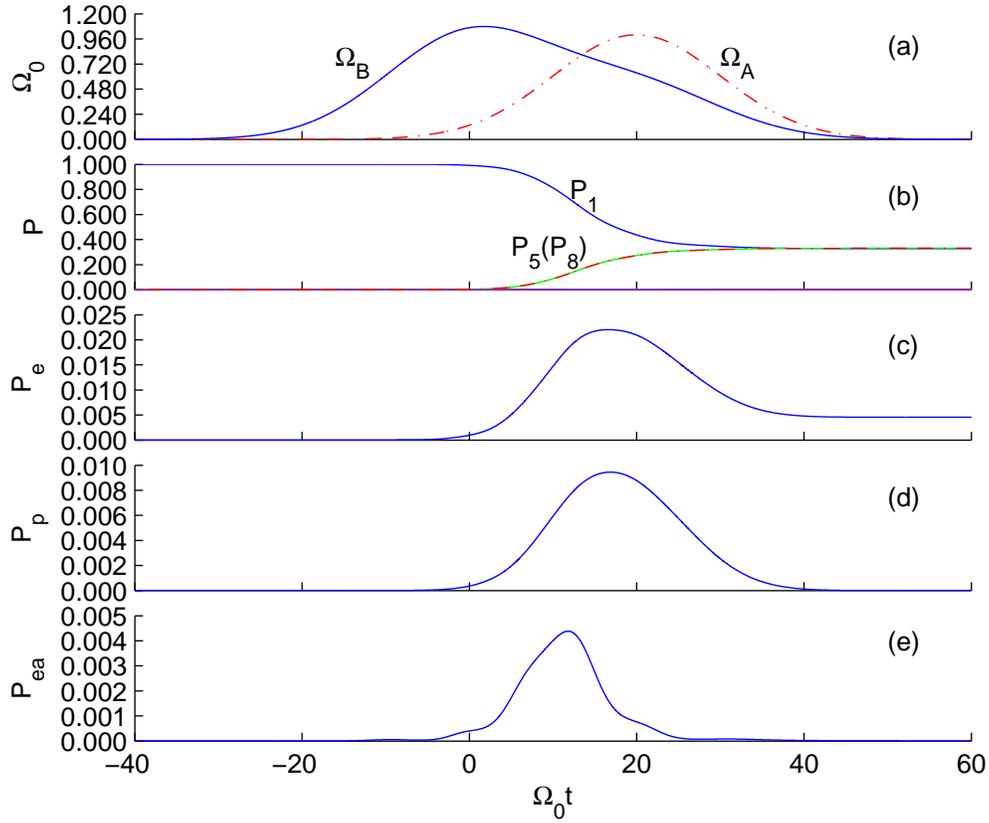}\caption{The simulation results of the two-qutrit entanglement generation
process, where we choose $g$, $\Omega_{0}$, $\tau$, $\kappa$, and
$\gamma$ the same as in Figure 2. Figure 3(a): the Rabi frequency of
$\Omega_{A}(t)$,
$\Omega_{B}(t)$. Figure 3(b): the time evolution of populations, where $P_{1}%
$, $P_{5}$, and $P_{8}$ mean the same as in Fig. 2. Figure 3(c):
error probability $P_{e}\left(  t\right)  $ defined by Eq. (8).
Figure 3(d): the probability $P_{p}$ with which one photon appears
in the cavity. Figure 3(e): the probability $P_{ea}$ of the atoms in
excited state $\left\vert
e_{i}\right\rangle $ $\left(  i=L\text{, }0\text{ ,}R\right)  $.}%
\end{figure}

\section{ Conclusion}

In summary, based on the STIRAP and f-STIRAP techniques, we have proposed two
schemes to generate entanglement of two $^{87}Rb$ atoms in a bi-mode cavity.
According to the simulation results, in the schemes for creation of atomic
entanglement states, the cavity mode, and the atomic excited states are nearly
not populated, so these schemes are hardly influenced by the atomic
spontaneous emission and the cavity decay. The error probability is also very
small during the process.

\textbf{Acknowledgments: }This work was funded by Natural Science Foundation
of China (Grant No 60677044).

\end{document}